# Reaction Engineering and Comparison of Electroenzymatic and Enzymatic ATP Regeneration Systems

Regine Siedentop,[a] Tobias Prenzel,[b] Siegfried R. Waldvogel,[b] Katrin Rosenthal,*[c] Stephan Lütz[a]

[a] Department of Biochemical and Chemical Engineering, TU Dortmund University, Emil-Figge-Str. 66, 44227 Dortmund (Germany)

[b] Department of Chemistry, Johannes Gutenberg-Universität Mainz, Duesbergweg 10–14, 55128 Mainz (Germany)

[c] School of Science, Constructor University, Campus Ring 1, 28759 Bremen (Germany)

**Correspondence:** Prof. Dr. Katrin Rosenthal (krosenthal@constructor.university). School of Science, Constructor University, Campus Ring 1, 28759 Bremen (Germany)

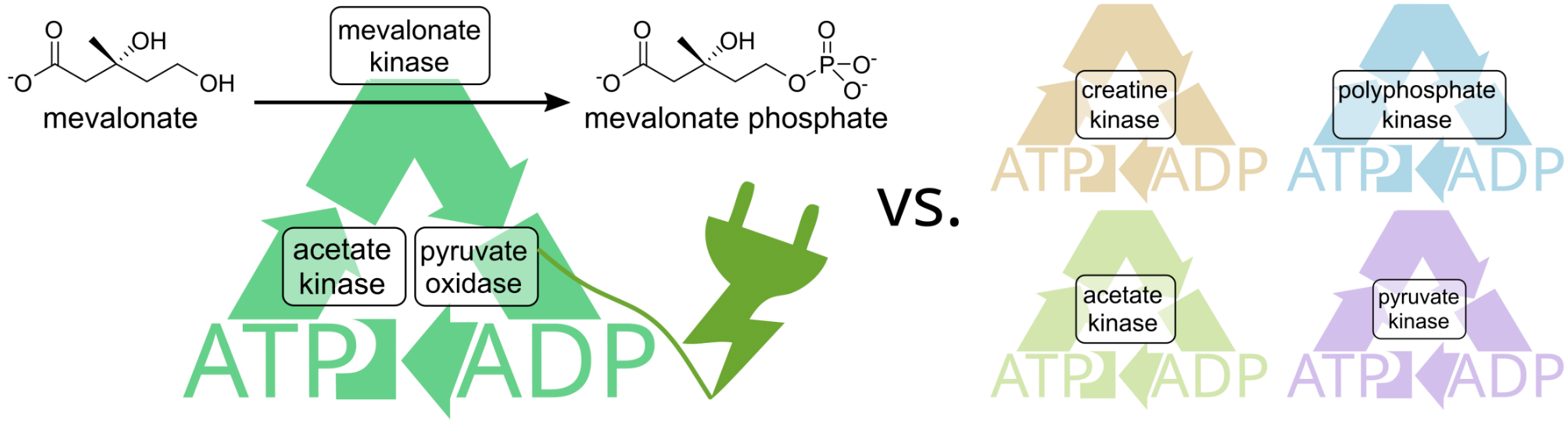

## Abstract

Adenosine-5'-triphosphate (ATP) plays a crucial role in many biocatalytic reactions and its regeneration can influence the performance of a related enzymatic reaction significantly. Here, we established an electrochemically coupled ATP regeneration by pyruvate oxidase and acetate kinase (ACK) for the phosphorylation of mevalonate catalyzed by mevalonate kinase. A yield of 84% for the product mevalonate phosphate was reached and a total turnover number for ADP of 68. These metrics are promising for the development of an economic feasible bioprocess and surpass many other enzymatic ATP regeneration systems. A comparison was made to polyphosphate kinases (PPKs), ACK, pyruvate kinase, and creatine kinase in terms of the phosphate donor properties and biocatalytic metrics of exemplary reactions. Furthermore, our system was expanded by a PPK that enables the phosphorylation of AMP, which can broaden the spectrum of applications even further.

## Introduction

Adenosine-5′-triphosphate (ATP) is a central cofactor of which many enzymes are dependent on [1]. The nucleotide functions as a carrier of chemical energy in form of high-energy phosphoanhydride bonds, that are released upon hydrolysis of the phosphate groups, or it serves as phosphate donor [2]. The cofactor is typically converted in equimolar amounts to the substrate and can play a crucial role in the activity and performance of the respective enzyme [3]. This is especially true for *in vitro* syntheses approaches, where no constant ATP supply by the metabolism is provided. Along with the possible inhibiting effects of ATP or its dephosphorylated counterparts, the nucleotide is cost intensive, which limits technical applications. A strategy to circumvent these challenges is the *in situ* production and regeneration of this cofactor [4]. This contributes also positively to the equilibrium of the reaction which is pulled towards the product side.

Various enzymatic systems have been reported to recycle ATP including polyphosphate kinases (PPKs) [5,6], acetate kinase (ACK) [7], pyruvate kinase (PK) [8], creatine kinase (CK) [9], or even by *in vitro* pathways mimicking glycolysis [10]. The *in vitro* ATP regeneration by ACK in combination with pyruvate oxidase (POX) has been introduced for cell-free protein synthesis (CFPS) and was recently implemented to a multi-step enzyme cascade [11,12]. The use of POX enables the recycling of inexpensive and stable inorganic phosphate as phosphate donor. In the reaction, pyruvate is decarboxylated and phosphorylated to acetyl phosphate, which in turn serves as phosphoryl donor for the ATP synthesis of ADP by ACK. During the POX-catalyzed reaction, two electrons are transferred to flavin adenine dinucleotide (FAD) as electron sink and then further transferred to $O_2$. Using oxygen as the final electron acceptor harbors some drawbacks such as the need for dissolved $O_2$ as well as the need for a catalase to prevent the accumulation of the reactive by-product $H_2O_2$, which is detrimental to enzyme stability and might evoke further side reactions. The flow of electrons in this reaction opens up the possibility of coupling to electrochemistry. Electrochemistry has already been applied to the regeneration of cofactors such as NAD(P)H or FAD [13–15] or other enzymatic redox reactions [16]. Electrochemical methods utilize electric current as a reagent by oxidation and reduction at an electrode [17,18]. They therefore do not require additional reagents, which can diminish the costs and the formation of by-products. Electrons can be transferred directly to the electrode or indirectly by utilizing a mediator, albeit direct transfer is rarely used [19]. ATP has already been shown to be indirectly regenerated electro-enzymatically either by an hydrogenase and ATP-synthase or by utilizing ACK and POX by transferring the electrons indirectly to an anode instead of oxygen, circumventing the mentioned drawbacks [20,21].

This work demonstrates the phosphorylation of ADP to ATP via coupling to an electrochemical reaction. The ATP regeneration system was subsequently applied for the enzymatic synthesis of mevalonate phosphate (MVAP), an important intermediate in the synthesis of isoprenoids. Furthermore, it was shown that an extension for the phosphorylation of AMP by PPKs is possible. Parameters to evaluate the reaction performance were calculated and compared to a variety of other ATP regeneration systems. In comparison, our approach performed well and is close to exceed thresholds for economic feasible bioprocess metrics such as total turnover numbers of enzymes and cofactor ($TTN_E$ and $TTN_{cofactor}$).

## Results and Discussion

### Electroenzymatic ATP Regeneration for Mevalonate Kinase

The phosphorylation of mevalonate (MVA) to MVAP plays an important role in the mevalonate pathway [22]. The phosphorylation is catalyzed by the enzyme mevalonate kinase (MVK) with ATP as phosphate donor (Figure 1). In the *in vitro* synthesis of the isoprenoid farnesyl pyrophosphate (FPP) by an enzyme cascade that includes MVK, we were able to show that ATP regeneration has a positive effect on the overall production performance of the enzyme cascade in terms of product titer [23]. The establishment of ATP regeneration by POX and ACK gives the advantages of the recycling of phosphate, a by-product of the FPP-production and further feeding of the FPP-substrate acetate, which is a by-product of POX. Furthermore, a high control of the reaction and no need of oxygen and catalase are given by the coupling to an electroenzymatic approach. Therefore, ACK and POX were chosen to implement an electroenzymatic regeneration of ATP for the reaction of MVK (Figure 1).

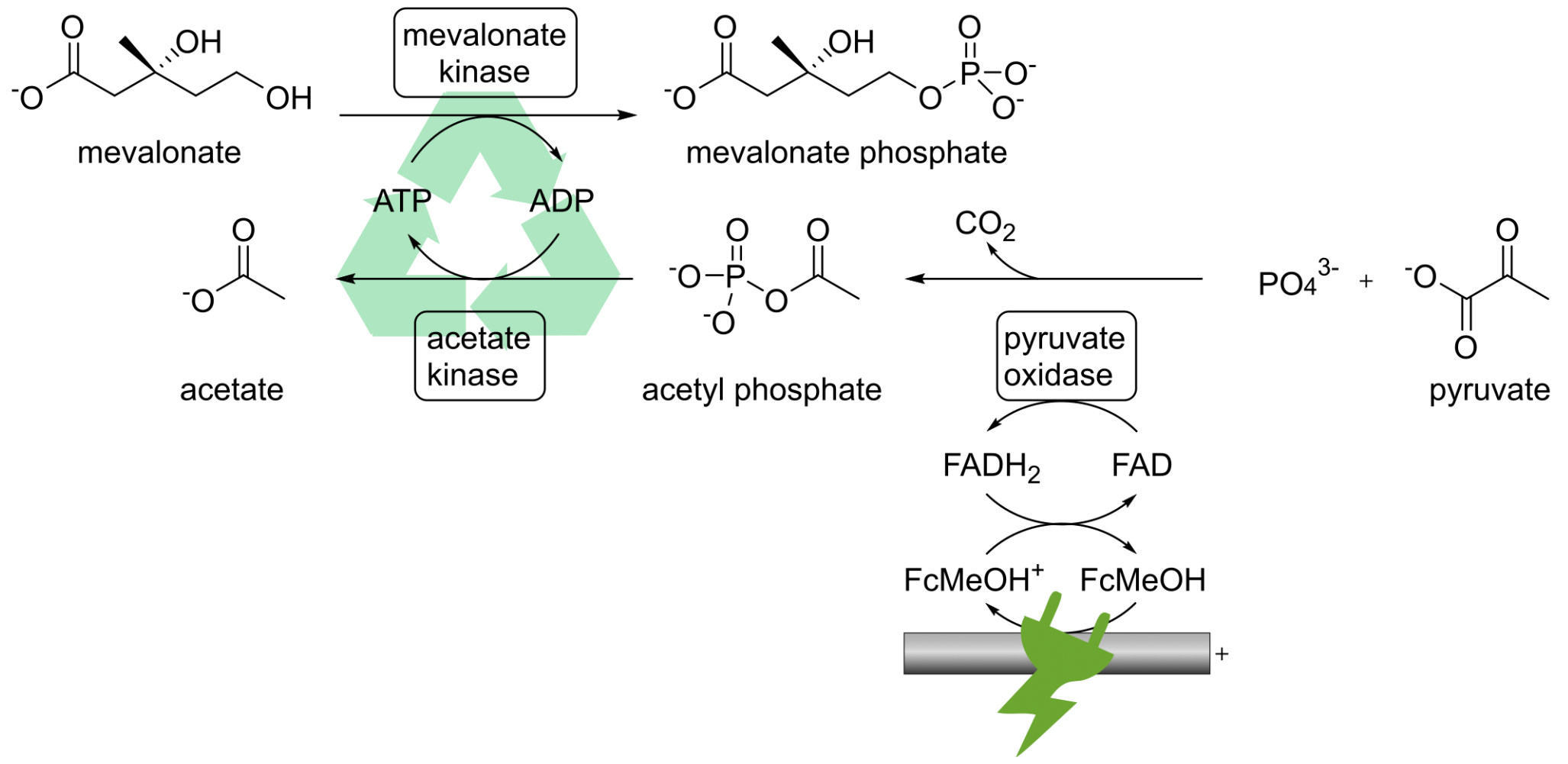

Figure 1: Scheme for the electroenzymatically coupled regeneration of ATP utilized by acetate kinase and pyruvate oxidase. ATP is consumed by mevalonate kinase for the reaction of mevalonate to mevalonate phosphate.

In this system, POX catalyzes the phosphorylation of pyruvate to acetyl phosphate with the elimination of $CO_2$ and release of two electrons, which are transferred to FAD, a cofactor serving as electron sink in the enzymatic reaction. Finally, the regeneration of FAD is achieved in a mediated anodic oxidation in which ferrocenyl-methanol (FcMeOH) is used as redox mediator for FAD-dependent enzymes. The oxidation of FAD is carried out by the ferrocenium ion, which in turn can be re-oxidized at the anode [24]. The synthesized acetyl phosphate serves as phosphate donor for the phosphorylation of ADP by ACK.

To prove, that the mediator is necessary for the reaction and to investigate if the reaction components have an influence on the standard potential, cyclic voltammetry (CV) was performed (Figure 2). As seen in Figure 2 B, without FcMeOH, no redox reaction takes place. As soon as the mediator was added, a standard potential $E^0$ of 0.21 V could be determined (Figure 2 A). This potential did not change by the addition of other components required for the ACK or MVK reaction (Figure 2 C). A previous report measured a similar standard potential of $E^0 = 0.22$ V for the mediator FcMeOH [21].

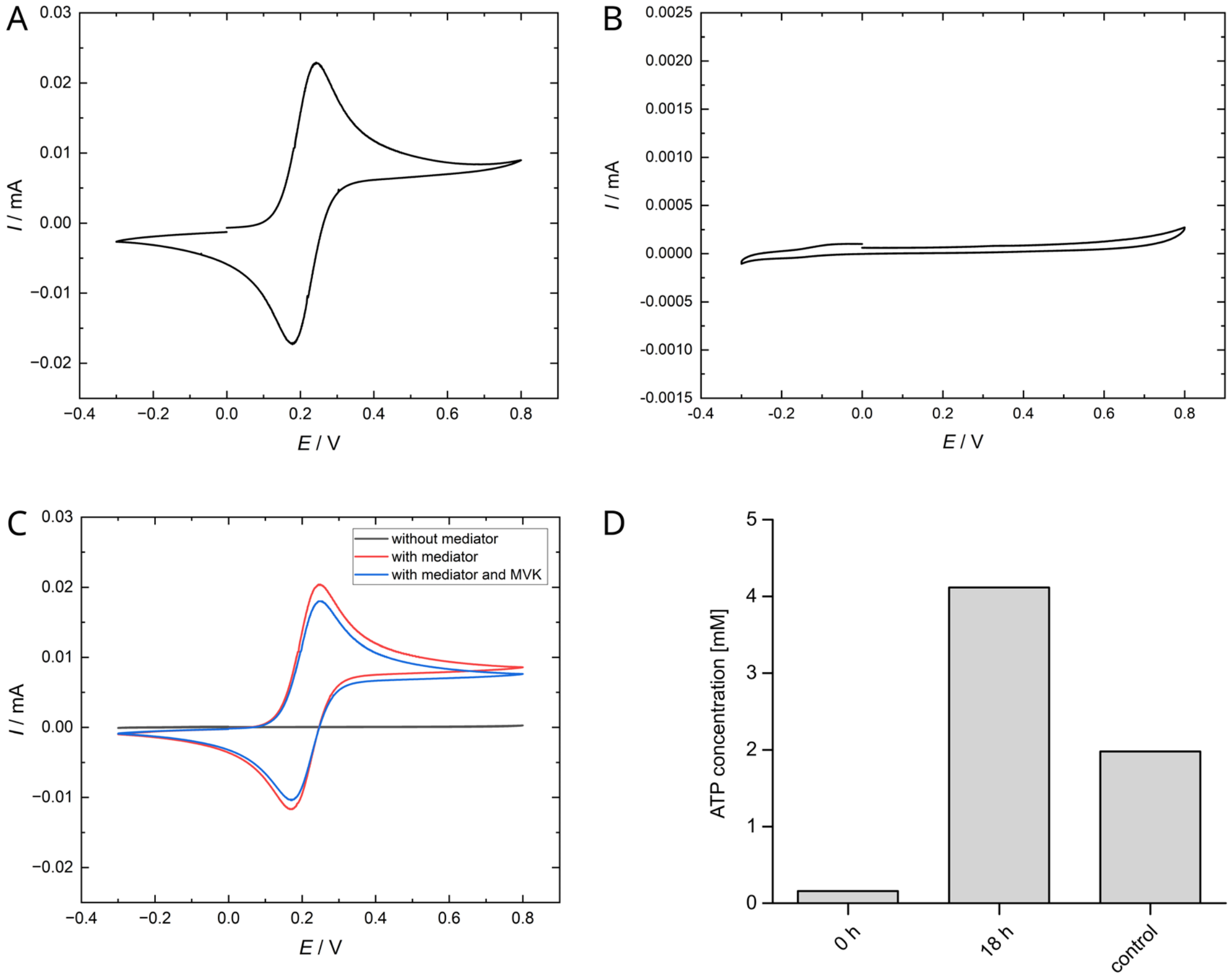

Figure 2: Cyclic voltammograms (working electrode: glassy carbon; counter electrode: platinum; reference electrode: Ag/AgCl in 3 M NaCl; scan rate: 16 mV/s) with various reaction components: (A) 5.7 mM FcMeOH in buffer (100 mM of Tris-HCl, 150 mM of NaCl, 10% glycerol, 20 mM of $MgCl_2$, pH 7.5), (B) 8.0 U $mL^{-1}$ POX, 0.2 mM DTT, 0.4 mM ThPP, 185.1 mM $PO_4^{3-}$, 18.5 mM $MgCl_2$, 185.1 mM pyruvate, 0.4 mM FAD in buffer, (C) 4.5 U $mL^{-1}$ POX, 0.1 mM DTT, 0.2 mM ThPP, 105.3 mM $PO_4^{3-}$, 10.5 mM $MgCl_2$, 105.3 mM pyruvate, 0.2 mM FAD, 3.7 mM FcMeOH, 1.1 mM ADP, 6.3 U $mL^{-1}$ ACK in buffer (red), 4.1 U $mL^{-1}$ POX, 0.1 mM DTT, 0.2 mM ThPP, 95.0 mM $PO_4^{3-}$, 9.5 mM $MgCl_2$, 95.0 mM pyruvate, 0.2 mM FAD, 3.3 mM FcMeOH, 1.0 mM ADP, 5.7 U $mL^{-1}$ ACK, 4.8 mM MVA, 190.0 mg $L^{-1}$ MVK in buffer (blue) and 8.0 U $mL^{-1}$ POX, 0.2 mM DTT, 0.4 mM ThPP, 185.1 mM $PO_4^{3-}$, 18.5 mM $MgCl_2$, 185.1 mM pyruvate, 0.4 mM FAD in buffer (grey). The standard potentials are 0.21 V. (D) ATP concentrations of electrochemically coupled ATP production (200 mM $Na_2HPO_4$, 200 mM pyruvate, 0.2 mM FAD, 0.2 mM ThPP, 3.5 mM FcMeOH, 0.1 mM DTT, 10 mM $MgCl_2$, 15 mM ADP, 4.3 U $mL^{-1}$ POX, 6 U $mL^{-1}$ ACK, filled up to 1.5 mL with buffer, stirred at 30 °C applying 0.086 mA $cm^{-2}$ for 18 h). A control was performed without current.

With the observation of a successful redox reaction at the anode using the mediator FcMeOH, the establishment of the ATP producing reaction was addressed. Therefore, a constant current approach was chosen in a simple undivided cell in a two-electrode setup. To ensure, that the electrons are not transferred to oxygen, dithiotreitol (DTT) is added to the reaction. ATP was produced with a yield of 52% corresponding to 4.12 mM ATP (Figure 2 D). As control, an assay was performed without current in which, as a result, less ATP with 25% yield was formed. These results are in agreement with a previous report of electrochemically coupled ATP regeneration with 67% yield at pH 7 and 21% conversion without applied current [21].

With the established ATP producing electroenzymatic cascade, a regeneration was anticipated for the phosphorylation reaction of MVK. Several conditions were tested, such as varying ADP or POX concentrations and different current densities between 0.043 mA $cm^{-2}$ and 0.862 mA $cm^{-2}$ (Table 1). Using current densities of 0.431 mA $cm^{-2}$ and higher led to gas formation at the electrodes and product yields remained low. Setting the current density low at 0.043 mA $cm^{-2}$ resulted in the prolongation of the reaction time to 24 h, which led to the precipitation of reaction components. The precipitant was hypothesized to be POX, since the assay was performed at pH 7.5 and the enzyme has an optimal pH below 7 [25]. Reaction conditions such as the pH were chosen for a possible implementation in the FPP-producing cascade [23].Nevertheless, a yield of 62% MVAP could be achieved if 1 mM ADP was present. Adding 0.1 mM ADP resulted in a reduced yield of 41% MVAP, but significantly higher $TTN_{ADP}$ of 61 compared to 9. In order to avoid POX precipitation, which was assumed to be time-dependent, the POX amount was increased, and the reaction time shortened by increasing the current density to 0.086 mA $cm^{-2}$. The yield could be increased to 84%, which corresponds to a $TTN_{ADP}$ of 68. Control experiments without applied current did not exceed a yield of 7% MVAP in any of the conditions.

Table 1: Results of various mevalonate phosphate (MVAP) producing assays utilizing electroenzymatically coupled ATP regeneration. If not mentioned otherwise, the assay composition was 200 mM $PO_4^{3-}$, 200 mM pyruvate, 0.2 mM FAD, 0.2 mM ThPP, 3.5 mM FcMeOH, 0.1 mM DTT, 10 mM $MgCl_2$, 0.1 mM ADP, 15 mM MVA, 4.3 U $mL^{-1}$ POX, 6 U $mL^{-1}$ ACK, 200 mg $L^{-1}$ MVK in 1.5 mL buffer (100 mM of Tris-HCl, 150 mM of NaCl, 10% glycerol, 20 mM of $MgCl_2$, pH 7.5). The reaction was stirred at 30 °C. Yields were calculated for the measured MVA concentration at the starting time of the reaction. TTN: total turnover number.

| Assay composition | Galvanostatic conditions | | Results | | |
|---|---|---|---|---|---|
| | Current density [mA $cm^{-2}$] | Reaction time | MVAP concentration [mM] | Yield [%] | $TTN_{cofactor}$ (mole product per mole cofactor) |
| 1 mM ADP | 0.043 | 29 h 5 min 10 s | 9.7 | 62 | $TTN_{ADP}$ = 9 |
| | 0.043 | 24 h 5 min 10 s | 6.1 | 41 | $TTN_{ADP}$ = 61 |
| | 0.431 | 5 h | 1.6 | 14 | $TTN_{ADP}$ = 16 |
| | 0.862 | 2.5 h | 0.7 | 7 | $TTN_{ADP}$ = 6 |
| 20 U $mL^{-1}$ POX | 0.086 | 18 h | 6.8 | 84 | $TTN_{ADP}$ = 68 |
| 0.1 mM AMP instead of ADP<br>20 mg $L^{-1}$ *Aj*PPK2<br>30 mM polyP | 0.086 | 18 h | 3.7 | 61 | $TTN_{AMP}$ = 37 |

For the implementation of this ATP regeneration system in enzymatic reactions with AMP instead of ADP as dephosporylated counterpart, a system with a PPK was tested. *Aj*PPK2 catalyzes the reaction of AMP to ADP with polyphosphate (polyP) as phosphate donor and was previously used in a similar reaction [26]. Using AMP as starting nucleotide, a MVAP yield of 61% was reached, which corresponds to a $TTN_{AMP}$ of 37. Albeit the lower yield, this result shows the compatibility of PPKs in electrochemically coupled reactions and that ATP regeneration can be extended to AMP producing reactions. The reaction of MVK utilizing two PPKs for ATP production and regeneration starting from AMP reached full conversion after the AMP and enzyme concentrations were optimized [26]. This shows that adjusting the substrate and enzyme concentrations for the POX and ACK system could also increase the yield and complete conversion could be reached.

**Evaluation of ATP Regenerating Systems**

Coupling electrochemistry to enzymes combines advantages such as substrate-, regio-, and stereoselectivity with a highly controllable reaction. Here, additional electron acceptors such as oxygen and the enzyme catalase can be omitted, since an external power source provides the electron sink for the reaction. Next to the ATP regeneration system used in this study, several other non-electrochemical ATP regeneration systems were developed such as utilizing the enzymes ACK, PK, CK, or PPK (Figure 3). An evaluation of the electrochemically coupled ATP regeneration by POX and ACK to the established systems is conducted here in terms of the properties of the phosphate donors and metrics of exemplary reactions such as TTN, turnover frequency (TOF), or space time yield (STY).

In order to regenerate ATP from adenosine phosphates *in vitro*, substrate level phosphorylation or transphosphorylation is mainly used. Therefore, another molecule as phosphate donor is necessary. To be suitable as phosphorylating agent, the Gibbs energy $\Delta G^0$ of the hydrolysis of the donor has to be larger than that of ATP, which is -30.5 kJ $mol^{-1}$ under standard conditions. For acetyl phosphate as substrate for ACK, creatine phosphate as substrate for CK, phosphoenolpyruvate (PEP) as substrate for PK, and polyP as substrate for PPK, (Figure 3) the Gibbs energy is between -30.5 kJ $mol^{-1}$ and -61.9 kJ $mol^{-1}$, making them suitable as ATP regenerating agents (Table 2).

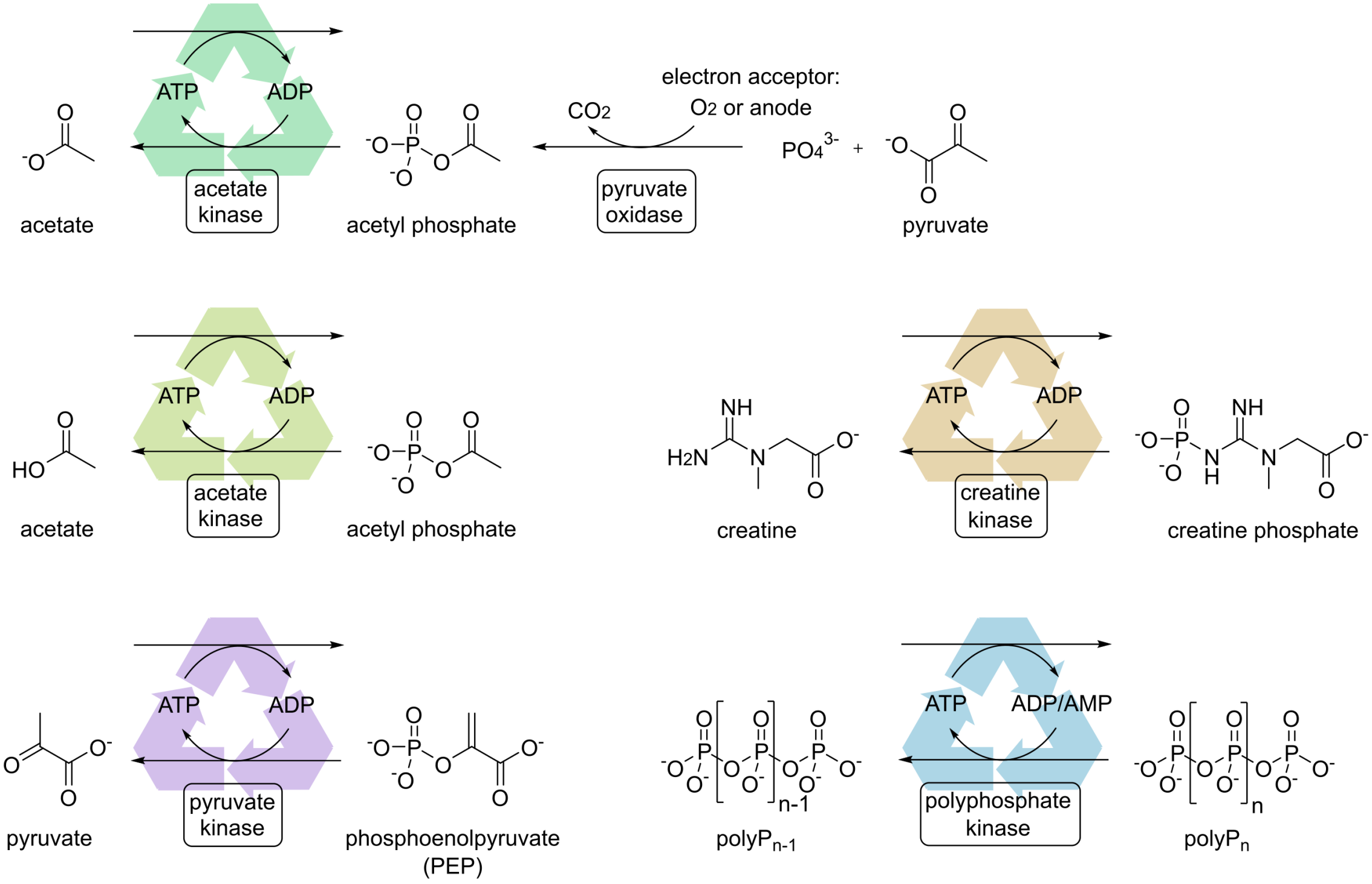

Figure 3: ATP Regeneration systems utilizing various enzymes and phosphate donors.

Also, the stability and hence, half-life of the phosphate donor plays an important role. The mentioned molecules are stable between a few hours to days and even years with polyP as the most stable molecule, of the selected phosphate donors. Acetyl phosphate has the shortest half-life of a few hours but is still more stable than ATP. Especially for large-scale applications, the purchasing costs of the phosphate donors are a valid argument to use or not use a regeneration system. The costs can vary significantly between a few cents as for polyP to thousands of euros per mole as for PEP. Pyruvate plus phosphate are the second most inexpensive substrates and especially the difference to acetyl phosphate is significant, emphasizing the role of acetyl phosphate-*in situ* production as in this work.

Table 2: Properties of the phosphate donors.

| Regenerating system | EC number of enzyme | Gibbs Energy $\Delta G^0$ of the phosphate donor [kJ $mol^{-1}$] | Half-life of phosphate donor [days] | Cost of phosphate donor [€ $mol^{-1}$][1] |
|---|---|---|---|---|
| ATP | - | -30.5 [28] | 0.25[2] [29,30] | 331.79 (0.5 kg pkg) |
| Pyruvate, $PO_4^{3-}$ – POX, ACK | 1.2.3.3, 2.7.2.1 | -43.3 (acetyl phosphate) | | 2.06 (pyruvate + $PO_4^{3-}$) (1 kg pkgs) |
| acetyl phosphate – ACK | 2.7.2.1 | -43.3 [31] | 0.34 [9,32] | 1855.32 (5 g pkg) |
| creatine phosphate – CK | 2.7.3.2 | -43.1 [28] | 12 [9] | 450.14 (25 g pkg) |
| PEP – PK | 2.7.1.40 | -61.9 [28] | 16-98 [9,32] | 7049.65 (1 g pkg) |
| polyP – PPK | 2.7.4.1 | -30.5 [33] | > 700[3] [34] | 0.48 (1 kg pkg) |

[1] package size in brackets [2] pH 8.4, 95 °C; [3] calculated with kH, first order reaction.

PolyP is the most stable and inexpensive phosphate donor, however the difference of the Gibbs energy for polyP to ATP is zero, which can make it difficult to drive thermodynamic unfavorable reactions forward [35]. On the other hand, PEP has the biggest difference to ATP in Gibbs energy, but it is also the most expensive phosphate donor. The driving forces of acetyl phosphate and creatine phosphate are similar, but they vary in their stability and their costs. Both drawbacks for acetyl phosphate can be overcome by its *in situ* production as shown in this work.

Further metrics to evaluate a reaction with included ATP regeneration are yield, conversion, TTN of enzymes ($TTN_E$, moles of product per mole of enzyme), TOF ($TTN_E$ per reaction time), TTN of the cofactor ($TTN_{cofactor}$, moles of product per mole of cofactor), and STY (grams of product per reaction volume and time). In table 3, various enzymatic systems with ATP regeneration are listed including the mentioned metrics. In order to be an economic feasible bioprocess, certain requirements for $TTN_E$, $TTN_{cofactor}$, productivity and yield have to be met. Threshold values for good biocatalyst performance are $TTN_E > 10^4\text{-}10^5$, TOF > 1 $s^{-1}$, $TTN_{cofactor} > 10^2\text{-}10^5$, and STY > 2-20 g $L^{-1}$ $h^{-1}$, though they have to be adjusted for the type of product such as pharma, fine, or bulk chemical [4,36–39].

In our work using the ATP regeneration system, MVAP was produced with a yield of 84% MVAP and with a production rate of 0.09 g $L^{-1}$ $h^{-1}$. The $TTN_{MVK}$ was 1218 with a $TOF_{MVK}$ of 0.02 $s^{-1}$ and the $TOF_{ACK}$ was 0.12 $s^{-1}$ (table 3). A $TTN_{ADP}$ of 69 was reached and the $TTN_{ACK}$ was 7489, which are close to the threshold values of an economic bioprocess. Hence, the electroenzymatic method holds great promise for an economical ATP regenerating process. Other studies used the same enzymes POX and ACK for ATP regeneration in 5-isobutyluridine synthesis with oxygen or an anode as electron acceptor [12,21]. Both reactions emphasize the potential of an ATP regeneration system by POX and ACK, since their metrics $TTN_{ACK}$, $TOF_{ACK}$, and $TTN_{ATP}$ are within the range for an economic feasible bioprocess. The electrochemically coupled ATP regeneration for 5-isobutyluridine synthesis admittedly has a lower $TTN_{ATP}$ with 94 than the oxygen coupled reaction with 145, but $TTN_E$ and TOF show better values ($TTN_{MTR\ kinase}$ of 3776 and $TTN_{ACK}$ of 75,334 vs $TTN_{MTR\ kinase}$ of 2925 and $TTN_{ACK}$ of 57,845, and $TOF_{ACK}$ 1.05 $s^{-1}$ vs 1.0 $s^{-1}$), indicating superiority of the current-driven reaction. During the synthesis of lacto-*N*-biose, ATP was recycled only one time by the POX, ACK and oxygen system, but higher $TTN_E$ for the ATP consuming enzyme galactokinase (GalK) with 49,800 and higher TOFs for GalK with 1.15 $s^{-1}$ and ACK with 1.68 $s^{-1}$ were reached [40].

Large-scale application of ACK on its own with added acetyl phosphate was proven to be applicable in a multi-kilogram scale and full conversion for glucose-6-phosphate (G6P)-production was reached [41]. Though, since acetyl phosphate has a low stability (Table 2), it is often added continuously during the reaction [42,43] and using this strategy, a $TTN_{ATP}$ of 109 could be reached for G6P production [42]. However, using acetyl phosphate as phosphate donor has some drawbacks such as cost or stability and an *in situ* production as in this work should be aspired (Table 2).

$TTN_{cofactor}$ of many regeneration systems have lower values than the system used in our work and even the extension to the regeneration starting from AMP shows comparable values to PPKs, which were successfully implemented in enzymatic syntheses either for AMP or ADP phosphorylation [44–46]. The threshold value for $TTN_{cofactor}$ of 100 is surpassed only from a few examples [9,12,42,47,48] and STY also seldomly reaches the threshold value of 2 g $L^{-1}$ $h^{-1}$ [12,49,50]. Our system is in the lower range of the production rates, which could be due to a low substrate loading. An increased substrate loading could also increase $TTN_E$ if the same amount of enzyme is used. Interestingly, $TTN_E$ for the regenerating enzyme is often higher than for the ATP consuming reaction. The values of $TTN_E$ reach the threshold value of $10^4$ more often than any other metric. This does not correlate with their TOFs, in which the reaction time is accounted for and only few examples surpass the threshold of 1 $s^{-1}$ [12,21,40,44,47]. Noticeably, often compromises have to be made for the metrics, as some exceed the values of e.g. $TTN_E$, but not for $TTN_{cofactor}$, TOF or productivity and vice versa. This stands in contrast with the POX and ACK system, which has examples with values above the threshold for all metrics.

To conclude, the POX and ACK regenerating system has advantages over other regenerating systems, such as the stability of the substrates or their low costs. The Gibbs energy of acetyl phosphate lies between all discussed phosphate donors and its *in situ* production overcomes the stability problem of acetyl phosphate and lowers the costs significantly, making it an attractive ATP regenerating system. Furthermore, in reactions with $PO_4^{3-}$ as by-product, it can be recycled by POX and lower therefore possible inhibitions by phosphate accumulation. Acetate, a by-product of POX might even serve as substrate for the enzymatic reaction [23,51]. The metrics for an economic bioprocess have promising values and exceed many of the examples with other ATP regenerating systems. Coupling the enzymatic system to electrochemistry, offers the advantages of a highly controlled reaction and the elimination of the need for oxygen and the enzyme catalase, making it an even more attractive regeneration method.

Table 3: Reactions with ATP regeneration systems and their corresponding metrics yield, conversion, total turnover number (TTN) for the ATP consuming and ATP regenerating enzymes and for the cofactor, as well as space time yield (STY).

| | Substrates and main product | Enzymes | Scale and reaction time | Yield | Conversion | $TTN_E$ (mole product per mole enzyme) | $TTN_{cofactor}$ (mole product per mole cofactor) | TOF ($TTN_E$ per reaction time) $[s^{-1}]$ | STY $[g\ L^{-1}\ h^{-1}]$ | Other | Ref. |
|---|---|---|---|---|---|---|---|---|---|---|---|
| ACK & POX | MVA + ADP + pyruvate + $PO_4^{3-}$ + current; MVAP | MVK, POX, ACK | 0.012 mmol MVA in 0.015 L; 18 h | 84% | n.d. | MVK: 1218<br>ACK: 7489 | 68 (ADP) | MVK: 0.02<br>ACK: 0.12 | 0.09 | - | This work |
| | MVA + AMP + pyruvate + $PO_4^{3-}$ + current; MVAP | MVK, POX, ACK, *Aj*PPK2 | 0.009 mmol MVA in 0.015 L; 18 h | 62% | n.d. | MVK: 657<br>ACK: 4039 | 37 (AMP) | MVK: 0.01<br>ACK: 0.06 | 0.05 | - | This work |
| | 5-isobutylribose + uracil + ATP + pyruvate + $PO_4^{3-}$ + current; 5-isobutyluridine | MTR kinase, UP, POX, ACK | 102 mmol 5-isobutylribose in 1 L flow reactor; 20 h | n.a. | 94% | MTR kinase: 3776;<br>ACK: 75,334 | 94 (ATP) | MTR kinase: 0.05;<br>ACK: 1.05 | 1.51 | Farradaic efficiency: 76% | [21] |
| | uracil + 5-isobutylribose + ATP + pyruvate + $PO_4^{3-}$ + $O_2$; 5-isobutyluridine | MTR kinase, UP, POX, ACK, catalase | 272 mmol uracil in 0.8 L; 16 h | 87% isolated yield | n.a. | MTR kinase: 2925;<br>ACK: 57,845 | 145 (ATP) | MTR kinase: 0.05;<br>ACK: 1.0 | 5.81 | - | [12] |
| | galactose + GlcNAc + ATP + pyruvate + $O_2$ ; Lacto-*N*-biose | GalK, LnbP, POX, ACK, catalase | 10 mmol galactose and 5 mmol GlcNAc in 0.1 L; 12 h | 96% | n.a. | GalK: 49,800;<br>ACK: 72,744 | 1 (ATP) | GalK: 1.15;<br>ACK: 1.68 | 1.53 | 0.96 $mol_{LNB}$ $mol_{GlcNAc}^{-1}$ | [40] |
| ACK | glucose + ATP + acetyl phosphate; G6P | Hexokinase, ACK | 1400 mmol glucose in 1.2 L, acetyl phosphate fed over 48h; 50h | 78% | n.a. | n.a. | 109 (ATP) | n.a. | 1.82 | - | [42] |
| | glucose + acetyl phosphate + ATP; G6P | GLK, ACK | 0.02 mmol glucose in 0.02 L; 1 h | n.a. | 97.4% | n.a. | 19 (ATP) | n.a. | 2.53 | - | [49] |
| | creatine + acetyl phosphate + ATP; creatine phosphate | creatine kinase, ACK | 427 mmol creatine in 3 L, acetyl phosphate fed over 36 h; 36 h | 55% | n.a. | n.a. | 47 (ATP) | n.a. | 0.08 | - | [43] |

| | | | | | | | | | | | | |
|---|---|---|---|---|---|---|---|---|---|---|---|---|
| CK | G1P + LacChr + creatine phosphate + UDP; α-GalChr | α 1,3-GalT, CK, GalU, PPA, GalE | 0.25 mmol G1P in 0.01 L; 49 h | 63% | n.a. | n.a. | 1575 (UDP) | n.a. | n.a. | - | [9] |
| PK | glucose + ATP + PEP; G6P | hexokinase, PK | 800 mmol glucose in 1.6 L; 8.5 d | 96% | n.a. | Hexokinase: $2*10^7$; PK: $4*10^7$ | 640 | Hexokinase: 27.23; PK: 54.47 | 0.61 | - | [47] |
| | arginine + ATP + PEP; arginine phosphate | arginine kinase, PK | 1200 mmol arginine in 3 L; 11.7 d | 75% | n.a. | n.a. | 20 (ATP) | n.a. | 0.0003 | - | [52] |
| | ribose-5-phosphate + $PO_4^{3-}$ + ATP + PEP; 5-phospho-D-ribosyl α-1-pyrophosphate (PRPP) | PRPP synthase, PK, adenylate kinase | 100 mmol ribose-5-phosphate in 1 L; 4 d | 75% | n.a. | n.a. | 3 (ATP) | n.a. | 0.30 | - | [53] |
| | fucose + lactose + GTP + ATP + PEP; 2'-fucosyllactose | FKP, FucT, PK | 0.61 mmol fucose in 0.01 L; 40 h | 91% | n.a. | FKP: 11,712; PK: 6853 | 4 (ATP) | FKP: 0.081; PK: 0.048 | 0.68 | - | [54] |
| | ATP + $SO_4^{2-}$ + PEP; 3'-phosphoadenosine-5'-phosphosulfate (PAPS) | ATP sulfurylase, APS kinase, pyrophosphatase, PK | 10 mmol ATP in 1 L; 6 h | n.a. | 98% | APS Kinase: 12,524; PK: 10,074 | 1 (ATP) | APS Kinase: 0.58; PK: 0.47 | 0.83 | - | [55] |
| PPK | 4-methoxybenzoic acid + glucose + $NADP^+$ + ATP + polyP 4-methoxybenzaldehyde + AMP | *Eb*PPK, CAR, GDH | 26 mmol 4-methoxybenzoic acid in 0.4 L; 22 h | 73.4% isolated yield | >99% | CAR: 735; *Eb*PPK: 2229 | 64 (ATP) | CAR: 0.01; *Eb*PPK: 0.03 | 0.30 | - | [46] |
| | CoA + acetate + AMP + polyp; acetyl-CoA | ACS, PAP, PPK, PPase | 0.004 mmol CoA in 0.4 mL; 18 h | 99.5% | 99.5% | ACS: 82,848; PPK: 38,861 | 40 (AMP) | ACS: 1.28; PPK: 0.6 | 0.45 | - | [44] |
| | cysteine + L-glutamic acid + glycine + ADP + polyp$_6$; GSH | GshF, PPK | 12.5 mmol cysteine in 0.25 L; 7 h | 80.24% | n.a. | GshF: 11,448; PPK: 9629 | 4 (ADP) | GshF: 0.45; PPK: 0.38 | 1.76 | - | [56] |
| | cysteine + L-glutamic acid + glycine + ADP + polyp; GSH | GshF, PPK | 0.035 mmol cysteine in 0.001 L; 5 h | 81.4% | n.a. | GshF: 55; PPK: 15 | 28 (ADP) | GshF: 0.0031; PPK: 0.00085 | 1.75 | Rate: 14.7 mM h$^{-1}$ | [57] |
| | alanine + Tyr + ATP + polyp$_6$; Ala-Tyr | Lal, PPK | 0.045 mmol alanine in 0.001 L; 3 h | 89.1% | n.a. | Lal: 2729 | 7 (ADP) | Lal: 0.25 | 3.37 | Rate: 13.4 mM h$^{-1}$ (25.6 mM h$^{-1}$ in first h) | [50] |
| | MVA + AMP + polyP; MVAP | MVK, *Aj*PPK2, *Sm*PPK2 | 0.0175 mmol MVA in 0.35 mL; 24 h | 99% | n.a. | MVK: 7850; *Sm*PPK2: 13,406 | 3 (AMP) | MVK: 0.09; *Sm*PPK2: 0.16 | 0.48 | - | [26] |
| | acetate + CoA + ATP + polyP; acetyl-CoA | *Ec*ACL, *Aj*PPK2, *Sm*PPK2 | 0.0001 mmol acetate in 0.1 mL; n.a. | n.a. | 30-35% | EcACL: 43-50; *Sm*PPK2: 222-259 | 300-350 (ATP) | n.a. | n.a. | - | [48] |
| | acetate + CoA + ATP + polyP; acetone | Acs, Thl, AtoDA, Adc, *Sm*PPK2, *Aj*PPK2 | 0.03 mmol acetate in 0.2 mL; 24 h | n.a. | 84% | Acs: 12,400; *Sm*PPK: 12,400 | 6 (ATP) | Acs: 0.14; *Sm*PPK: 0.14 | 0.15 | - | [51] |
| | D-fructose + ATP + polyP; D-allulose 1-phosphate | DPE, RhaB, *Te*PPK | 0.02 mmol D-fructose in 0.001 L; 6 h | n.a. | 99% | n.a. | 10 (ATP) | n.a. | 0.85 | - | [58] |

n.a. not available; n.d. not determined; ACK: acetate kinase; ACL: arylamine-*N*-acetyltransferase; ACS: acetyl-CoA synthase; ADC: acetoacetate decarboxylase; ADP: adenosine-5’-diphosphate; Ala: alanine; AMP: adenosine-5’-monophosphate; APS: adenosine-5‘-phosphosulfate; AtoDA: acetoacetyl-CoA transferase; ATP: adenosine-5’-triphosphate; CAR: carboxylic acid reductase; CoA: coenzyme A; CK: creatine kinase; DPE: D-psicose epimerase; FKP: L-fucokinase/GDP-L-fucose phosphorylase; FucT: α-1,2-fucosyltransferase; G1P: glucose-1-phosphate; G6P: glucose-6-phosphate; GalK: galactokinase; GalE: UDP-Gal 4‘-epimerase; GalU: glucose-1-phosphate uridylyltransferase; GDH: glucose dehydrogenase; GlcNAc: *N*-acetylglucosamine; GLK: glucokinase; GSH: Glutathione; GshF: GSH synthetase; LacChr: (4,5-dimethoxy-2-nitrophenyl)methyl O-(β-D-galactopyranosyl)-(1⟶4)-β-D-glucopyranoside; Lal: L-amino acid ligase; LnbP: lacto-*N*-biose phosphorylase; MTR kinase: S-methyl-5-thioribose kinase; MVA: mevalonate; MVAP mevalonate phosphate; MVK; mevalonate kinase; PAP: polyphosphate-AMP phosphotransferase ; PEP: phosphoenolpyruvate; PK: pyruvate kinase; PPA: inorganic phosphatase; PPK: polyphosphate kinase; polyP: polyphosphate; POX: pyruvate oxidase; RhaB: L-rhamnulose kinase; Thl: thiolase; Tyr: tyrosine; UDP: uridine-5‘-diphosphate; UP: uridine phorphorylase.

## Conclusion

An ATP regenerating system coupled to electrochemistry utilizing POX and ACK was established for the ATP consuming enzyme MVK reaching a yield of 84% for the product MVAP and a $TTN_{ADP}$ of 68. Furthermore, it could be shown, that the coupling of PPKs is possible for an AMP generating reaction, which broadens the spectrum of applications even more. In addition, its substrates have some advantages in comparison to other ATP regenerating systems, such as the costs or the stability, emphasizing the *in situ* production of acetyl phosphate. Further metrics of our reaction such as $TTN_{ATP}$ and $TTN_{ACK}$ show promising values for an economic feasible bioprocess and exceed many other examples. This indicates that even higher ATP concentrations and $TTN_{cofactor}$ are expected by the optimization and fine-tuning of the reaction or directed evolution of the enzymes. The electric source is a strong driving force towards the product side for the ATP-coupled reaction and many applications await for the testing for their compatibility.

## Materials and Methods

### Materials

Chemicals were purchased from Acros Organics (ThermoFisher Scientific, Waltham, MA, USA), AppliChem (AppliChem GmbH, Darmstadt, Germany), Merck (Merck KGaA, Darmstadt, Germany), Roth (Carl Roth, Karlsruhe, Germany), Santa Cruz Biotechnology (Santa Cruz Biotechnology, Inc., Dallas, TX, USA), TCI (TCI Deutschland GmbH, Eschborn, Germany), ThermoFisher (ThermoFisher Scientific, Waltham, MA, USA), and VWR (VWR international GmbH, Darmstadt, Germany).

### Protein production

Plasmids of MVK and *Aj*PPK2 were kindly provided by Frank Schulz and Jennifer Andexer and they were expressed and purified as explained in [26,59]. POX and ACK were purchased from Merck (Merck KGaA, Darmstadt, Germany).

### Cyclic Voltammograms

Electrochemical experiments were performed using a BioLogic SP-50e potentiostat with EC Lab® software (BioLogic, V11.43). Cyclic voltammograms were performed using a platin wire (57 mm, 0.5 mm) as counter electrode, an Ag/AgCl (3 M NaCl, RE-1BP) as reference electrode and a glassy carbon electrode (GCE, OD: 6 mm, ID 3 mm) as working electrode. The potential was varied between -0.3 V to 0.8 V. The concentrations of the components were 5.7 mM FcMeOH in buffer (100 mM of Tris-HCl, 150 mM of NaCl, 10% glycerol, 20 mM of $MgCl_2$, pH 7.5) (Figure 2 A), 8.0 U $mL^{-1}$ POX, 0.2 mM DTT, 0.4 mM ThPP, 185.1 mM $PO_4^{3-}$, 18.5 mM $MgCl_2$, 185.1 mM pyruvate, 0.4 mM FAD in buffer (Figure 2 B), 4.5 U $mL^{-1}$ POX, 0.1 mM DTT, 0.2 mM ThPP, 105.3 mM $PO_4^{3-}$, 10.5 mM $MgCl_2$, 105.3 mM pyruvate, 0.2 mM FAD, 3.7 mM FcMeOH, 1.1 mM ADP, 6.3 U $mL^{-1}$ ACK in buffer (Figure 2 C, red), 4.1 U $mL^{-1}$ POX, 0.1 mM DTT, 0.2 mM ThPP, 95.0 mM $PO_4^{3-}$, 9.5 mM $MgCl_2$, 95.0 mM pyruvate, 0.2 mM FAD, 3.3 mM FcMeOH, 1.0 mM ADP, 5.7 U $mL^{-1}$ ACK, 4.8 mM MVA, 190.0 mg $L^{-1}$ MVK in buffer (Figure 2 C, blue).

### Electrochemical ATP production and regeneration

The assays were performed in a HPLC vial, using a magnetic stirrer and a water bath, heated to 30 °C. Assay composition for ATP production was 200 mM $Na_2HPO_4$, 200 mM pyruvate, 0.2 mM FAD, 0.2 mM ThPP, 3.5 mM FcMeOH, 0.1 mM DTT, 10 mM $MgCl_2$, 15 mM ADP, 4.3 U $mL^{-1}$ POX, 6 U $mL^{-1}$ ACK, filled up to 1.5 mL with Tris-based buffer (100 mM of Tris-HCl, 150 mM of NaCl, 10% glycerol, 20 mM of $MgCl_2$, pH 7.5). A glassy carbon rod (2.0 mm, 1.16 $cm^2$) was used as anode and a platinum wire (0.5 mm, 0.28 $cm^2$) as a cathode. Constant current electrolysis was applying 0.086 mA $cm^{-2}$ for 18 h (6.48 C). For ATP regeneration, 0.1 mM ADP, 15 mM MVA, and 200 mg $L^{-1}$ MVK were added, if not mentioned otherwise with constant densities between 0.043 mA $cm^{-2}$ and 0.862 mA $cm^{-2}$ (4.32-9.00 C). Samples were inactivated at 95 °C for 5 min.

### Analytics

Samples were centrifuged and diluted with water prior to injection. MVA and MVAP were analyzed using a SeQuant® ZIC®-pHILIC 5 µm polymer, 150 × 2.1 mm column (Merck KGaA, Darmstadt, Germany) with a LC-MS (1260 Infinity II LC system combined with 6120 Quadrupole MS (Agilent, Santa Clara, CA, USA)). The column oven was heated to 40 °C, the injection volume was set to 3 µL and the flow rate was set to 0.2 mL $min^{-1}$. The following gradient of mobile phase A (90% 10 mM $NH_4Ac$, pH 9.2, 10% ACN) and mobile phase B (90% acetonitrile, 10% 10 mM $NH_4Ac$, pH 9.2) was used: 0 min: 0% A; 1 min: 0% A; 30 min: 75% A; 34 min: 100% A; 39 min: 100% A; 49 min: 0% A; 64 min: 0% A. The MS measurement was performed in negative mode. The parameters for electron spray ionization (ESI) were set to the following: drying gas temperature: 350 °C, drying gas flow: 12 L $min^{-1}$, nebulizer pressure: 35 psig, capillary voltage: 3500 V. Analytes were detected in the selected ion mode (SIM) with either m/z of 426, 426.1, and 426.2 for ADP and 506, 506.1 and 506.2 for ATP or m/z of 147.0 and 147.1 during the first 15 min to detect MVA and 226.9, 227, 227.1, and 455.0 for monomeric and dimeric MVAP from 15 to 33 min. Yields were calculated for the measured MVA concentration at the starting time before applying current to the reaction.

**Acknowledgments**

This research was funded by Deutsche Forschungsgemeinschaft (DFG) under the priority programme SPP 2240 “eBiotech” (Bioelectrochemical and engineering fundamentals to establish electro-biotechnology for biosynthesis–Power to value-added products) (WA 1276/27-1 and grant agreement No 445751305).

The authors would like to thank Frank Schulz for providing the plasmid of MVK and Jennifer Andexer for providing the plasmids of the PPKs. The authors would like to acknowledge the support with analytics by Sascha Nehring and Don Marvin Voss.